\documentclass[12pt]{article}
\usepackage{amsmath,amssymb,bm,graphicx}
\usepackage{color} 
\usepackage{hyperref}
\usepackage{slashed}
\usepackage{caption}
\usepackage{subcaption}

\newcommand{\tp}{\text{p}}

\newcommand{\sgn}{\text{sgn}}

\newcommand{\da}{{\downarrow }}
\newcommand{\ua}{{\uparrow }}
\usepackage{cite}
\numberwithin{equation}{section}

\begin{document}


\begin{center}
{\Large{\bf Neutrino Oscillations\\\vskip .2 truecm by a Manifestly Coherent Mechanism\\\vskip .2 truecm and Massless vs. Massive Neutrinos}}
\end{center}
\vskip .5 truecm
\begin{center}
{\bf {  Anca Tureanu}}
\end{center}

\begin{center}
\vspace*{0.4cm} 
{\it $^{a)}$Department of Physics, University of Helsinki,\\ P.O.Box 64, 
FI-00014 University of Helsinki,
Finland\\
\vskip 0.3cm
$^{b)}$Helsinki Institute of Physics,\\ P.O.Box 64, 
FI-00014 University of Helsinki,
Finland
}
\end{center}
\vspace*{0.2cm} 
\begin{abstract}
The neutrino oscillations in vacuum are derived in a manifestly coherent scheme. The mechanism is operative in a quantum field theoretical framework, justifying nevertheless a formal analogy with quantum mechanical two- (or more) level systems and their oscillatory behaviour. Both the flavour states and the massive states are eigenstates of certain Hamiltonians which, in special conditions, can be argued to share the same Hilbert space. In this scheme, flavour neutrinos are massless and play the role of asymptotic states for any interactions, including the weak interactions, while massive neutrinos are effective propagation states. The vacuum is interpreted as a medium, where the flavour neutrinos undergo coherent forward scatterings which modify their energy and mix their flavour.  The treatment of matter conversion and MSW effect fits in naturally; the extension to other neutral particle oscillations, like $K_0-\bar K_0$, is straightforward. The scheme is eclectic insofar as it combines seamlessly quantum field theory and quantum mechanics.
\end{abstract}

\section{Introduction}\label{intro}

The theory of neutrino oscillations in vacuum is still arousing controversies, yet the gist of it is generally agreed upon \cite{Bilenky_hist,mohapatra, fukugita, giunti, bilenky, xing, valle, roulet}. It can be summarized by saying that neutrinos are produced in weak interactions as coherent superpositions of spinor states of different masses. The free propagation in vacuum of the superposition of states is controlled by the mass difference squared, for ultrarelativistic neutrinos, leading to the oscillation of the flavour quantum number. As a result, there exists a certain probability to detect a neutrino of a different flavour compared to the one that was produced. The standard framework of neutrino oscillations is based on Pontecorvo's extension of the state mixing and oscillation paradigm from the $K_0-\bar K_0$ system \cite{GMP, PP} to neutrinos \cite{Pontecorvo_1, Pontecorvo_2, Gribov_Pontecorvo, Bilenky-Pontecorvo} (see also \cite{BP_review}).

For the interpretation of vacuum neutrino oscillations experiments, the following two assumptions are necessary:
\begin{enumerate}

\item {\it The production and detection of neutrinos by charged current weak interactions takes place with strict conservation of flavour quantum number.} This assumption reflects the fact that neutrinos by themselves are never observed, and their flavour is inferred from the type of charged lepton that accompanies them in the interaction. 

\item {\it The flavour oscillation happens only during the free propagation of neutrinos, due to the flavour mixing terms in the Lagrangian.} Those terms induce the flavour neutrino states to be coherent superpositions of different mass states.

\end{enumerate}

{These conditions are purely theoretical constructions -- it is impossible to experimentally confirm the strict separation of flavour-conserving weak interactions from flavour-violating propagation in vacuum. One can state that, to the present level of experimental precision, weak interactions conserve flavour, since "zero-distance conversion" of neutrinos has not been observed\footnote{Zero-distance conversion of neutrinos means that neutrinos produced in conjuction with electrons, for example (beta decay), have a probability to interact with muons, and this probability is independent on the length from the production point. The process is usually discussed in the context of mixing with heavy sterile neutrinos. Nevertheless, such conversions have not been observed.}. This, however, does not preclude the flavour violation in weak interactions.}

Below are summarized the main points of the traditional approach to neutrino oscillations.

It is generally believed that the neutrino interactions are described by the Standard Model Lagrangian, with additional mass terms, which also mix the flavour neutrino fields. In this paper we shall consider, for simplicity, a two-neutrino model with Dirac mass terms. The generalization to any number of flavours and the inclusion of Majorana terms is straightforward. With these assumptions, the Lagrangian of the two neutrino fields ${\nu_\ell}(x)$, with $\ell=e,\mu$,  reads:
\begin{eqnarray}\label{Lagr}
{\cal L}={\cal L}_0+{\cal L}_{mass}+{\cal L}_{CC}+{\cal L}_{NC},
\end{eqnarray}
where ${\cal L}_0$ contains the kinetic terms:
\begin{eqnarray}\label{L0}
{\cal L}_0=\sum_{\ell=e,\mu}\left[\bar{\nu}_{\ell L}(x)i\slashed{\partial}\nu_{\ell L}(x) +\bar{\nu}_{\ell R}(x)i\slashed{\partial}\nu_{\ell R}(x)\right],
\end{eqnarray}
${\cal L}_{mass}$ contains the mass and mixing terms\footnote{Throughout this paper we work with the convention that all the mixing in lepton sector is manifest in neutrino mixing.}:
\begin{eqnarray}\label{Lmass}
{\cal L}_{mass}=- \left(\begin{array}{c c}
            \bar{\nu}_{eL}(x)&  \bar{\nu}_{\mu L}(x)
            \end{array}\right) \left(\begin{array}{c c}
            m_{ee} &m_{e\mu}\\
            m_{e\mu}&m_{\mu\mu}
            \end{array}\right)\left(\begin{array}{c}
            \nu_{e R}(x)\\
            {\nu_{\mu R}}(x)
            \end{array}\right)+h.c.,
\end{eqnarray}
${\cal L}_{CC}$ describes the charged current interactions:
\begin{eqnarray}\label{LCC}
{\cal L}_{CC}=-\frac{g}{\sqrt2}\left[\bar{\nu}_{eL}(x)\gamma_{\mu} e_L(x)+\bar{\nu}_{\mu L}(x)\gamma_{\mu} \mu_L(x)\right]W^\mu(x)+h.c.,
\end{eqnarray}
and ${\cal L}_{NC}$ describes the neutral current interactions.
Although the right-chiral fields are sterile for weak interactions, we still denote them by a flavour index, to keep a certain order in the formulas.

Written in terms of flavour field operators as above, the Lagrangian is invariant under the global $U(1)$ flavour symmetry, with the exception of the terms proportional to the mixing parameters $m_{e\mu}$.

The bilinear terms in \eqref{Lmass} mix the two {\it flavour field operators} and the Lagrangian is diagonalized by a change of variables, using the orthogonal transformation\footnote{We consider $m_{ee}m_{\mu\mu}>m_{e\mu}^2$ in order to obtain positive mass eigenstates in \eqref{masses} with a simple rotation like \eqref{rotation}. If this condition is not fulfilled, one tweaks a bit the rotation matrix transforming it into a unitary matrix and then apply the Autonne--Takagi procedure, which will necessarily lead to real and positive masses. The procedure is well known and can be found in any neutrino physics book. Here we prefer not to clutter unnecesarily the formulas.}:
\begin{eqnarray}\label{rotation}
\left(\begin{array}{c}
            {\nu_e}(x)\\
            {\nu_{\mu}}(x)
            \end{array}\right)= \left(\begin{array}{c c}
            \cos\theta &\sin\theta\\
           -\sin\theta&\cos\theta
            \end{array}\right)\left(\begin{array}{c}
            \nu_{1}(x)\\
            \nu_{2}(x)
            \end{array}\right),
\end{eqnarray}
with 
\begin{equation}\label{theta}
\tan2\theta=\frac{2m_{e\mu}}{m_{\mu\mu}-m_{ee}}
\end{equation}
and
\begin{eqnarray}\label{masses}
m_{1,2}&=&\frac{1}{2}\left\{\left(m_{ee}+m_{\mu\mu}\right)\mp\sqrt{(m_{ee}^2-m_{\mu\mu}^2)^2+4m^2_{e\mu}}\right\}, \ \ \ \ m_{ee}m_{\mu\mu}>m_{e\mu}^2.
\end{eqnarray}
In terms of the new field variables $\nu_{i}(x),\ i=1,2$, the Lagrangian terms read:
\begin{eqnarray}\label{Ldiag}
{\cal L}_0+{\cal L}_{mass}&=&\bar\nu_1(i\slashed{\partial}-m_1)\nu_1 +\bar\nu_2(i\slashed{\partial}-m_2)\nu_2
\end{eqnarray}
and
\begin{eqnarray}\label{LCC'}
{\cal L}_{CC}&=&-\frac{g}{\sqrt2}\Big[\cos\theta\,\bar\nu_{1L}(x)\gamma_{\mu} e_L(x)+\sin\theta\,\bar\nu_{2L}(x)\gamma_{\mu} e_L(x)\cr
&-&\sin\theta\,\bar\nu_{1L}(x)\gamma_{\mu L} \mu(x)+\cos\theta\,\bar\nu_{2L}(x)\gamma_{\mu} \mu_L(x)\Big]W^\mu
+h.c.
\end{eqnarray}
In this form, the effect of the mixing terms is dispersed in the whole Lagrangian, and the flavour $U(1)$ symmetry is manifestly broken in the weak interaction terms, in contrast to the expression \eqref{LCC}.

In order to fulfill the first requirement above (i.e. flavour conservation in the weak interactions of leptons), it is customary to consider that the weak interactions are described by the Lagrangian \eqref{LCC} and not by \eqref{LCC'}. On the other hand, the asymptotic states are considered to be the massive neutrino states of masses $m_{1,2}$, and not the flavour states associated to $\nu_e(x),\nu_\mu(x)$. In short, the weak interactions should produce and annihilate massive states in certain prescribed/coherent combinations, thus violating the principles of QFT (more about it in Sect. \ref{superposition})\footnote{In Appendix \ref{app_flavour_violation} it is shown that this alteration of the basic rules of quantum field theory actually does not help justify the conservation of flavour in the processes involving neutrinos, not even at high energies (see also \cite{GKL}).}. {Finding ways out of this predicament is a constant preocupation in neutrino physics up to these days, which sparkles still controversies (for a limited selection of references, see \cite{GKL, AS-coher, Jones, Evslin, Kupiainen, Fujikawa_PI, Tureanu, Cozzella, Ho, Ochman, Giunti:2003dg, Kayser, Kiers, Nussinov}).

To describe the vacuum oscillations, one veers abruptly from quantum field theory to the quantum mechanics of two-level systems. The procedure hinges on the {\it ad hoc} assumption that the flavour states produced by weak interactions, without being quanta of the flavour field operators ${\nu_\ell}(x)$, are however coherent superpositions of massive Fock states. In Sect. \ref{superposition} we explain why this assumption collides with (and succumbs to) the basic principles of quantum field theory.}

In short, for a mixing of two neutrino flavours, the standard approach is to {\it postulate} the existence of the flavour states $|{\nu_e}\rangle$ and $|{\nu_\mu}\rangle$, defined by:
\begin{eqnarray}\label{states_mix}
\left(\begin{array}{c}
            |{\nu_e}\rangle\\
            |{\nu_{\mu}}\rangle
            \end{array}\right)= \left(\begin{array}{c c}
            \cos\theta &\sin\theta\\
           -\sin\theta&\cos\theta
            \end{array}\right)\left(\begin{array}{c}
            |\nu_1\rangle\\
            |\nu_2\rangle
            \end{array}\right),
\end{eqnarray}
where $|\nu_1\rangle$ and $|\nu_2\rangle$ represent the massive neutrino Fock states, with the masses $m_1$ and $m_2$, while $\theta$ is the mixing angle \eqref{theta}. The coherence of the superposition is presumed without justification (though this fact is glossed over and hardly ever explicitly acknowledged in the literature) and it is encoded in the fact that the mixing matrix is strictly the rotation matrix in \eqref{states_mix} and no random phase is allowed. Although in the standard approach the massive states are characterized by the same momentum\footnote{This assumption has been amply debated in the literature.}, this is not implicit in formula \eqref{states_mix}. Curiously, formula \eqref{states_mix} is believed to express a well-defined superposition of {\it undefined} massive states.

It is also standard to postulate that the massive neutrinos satisfy the non-relativistic Schr\"odinger equations:
\begin{eqnarray}\label{Sch_eq}
i\frac{\partial}{\partial t}\left(\begin{array}{c}
            |{\nu_1}(\bf p)\rangle\\
            |{\nu_{2}(\bf p)}\rangle
            \end{array}\right)= \left(\begin{array}{c c}
            E_1 &0\\
           0&E_2
            \end{array}\right)\left(\begin{array}{c}
            |{\nu_1}(\bf p)\rangle\\
            |{\nu_{2}(\bf p)}\rangle
            \end{array}\right),\ 
\end{eqnarray}
with
\begin{equation}
E_i=\sqrt{{\bf p}^2+m_i^2}=|{\bf p}|\left(1+\frac{m_i^2}{2|{\bf p}|^2}+{\cal O}\left(\frac{m_i^4}{|{\bf p}|^4}\right)\right), \ \ \ i=1,2.
\end{equation}
The standard oscillation probability, in the approximation that neutrinos are ultrarelativistic, is 
\begin{eqnarray}\label{prob_approx}
P_{\nu_e\to\nu_\mu}=&\sin^22\theta\sin^2\left(\frac{\Delta m^2}{4E}L\right), \ \ \ \Delta m^2=m_2^2-m_1^2,
\end{eqnarray}
where $E$ is the energy of the neutrinos in the beam and $L$ is the distance between the neutrino production and detection points \cite{mohapatra, fukugita, giunti, bilenky, xing, valle, roulet}, taking into account that in the limit where the speed of the neutrinos is almost the speed of light, we can approximate the time of flight by the distance traveled ($t\sim L$).

In this paper, we shall show that, in certain conditions, formula \eqref{states_mix} can be derived for states of identical momenta, when $m_1,m_2\ll \tp$. In Sect. \ref{superposition} we argue why this formula cannot be postulated in general and present its conceptual drawbacks. In Sect. \ref{coh sch} we present the derivation of formula \eqref{states_mix} in a quantum field theoretical scheme which justifies a formal quantum mechanical treatment of the oscillations. In the same section we discuss the oscillations and conversion in matter. In Sect. \ref{flavour_violation} we evaluate the magnitude of the flavour violation in the weak interactions of neutrinos and show that it is below the present experimental capabilities of detection, and in Sect. \ref{kaons} we discuss comparatively the neutral kaon oscillations and how the present scheme fits into the description of mixing proposed in the original papers of Gell-Mann, Pais and Piccioni \cite{GMP, PP}.

\section{Superposition of quantum states in standard approach to oscillations}\label{superposition}

The standard approach with all its variations relies on the assumption that the asymptotic/free neutrino states are the Fock states of the massive spinor fields $\nu_1(x), \nu_2(x)$. Nevertheless, for some reason that is never explained, those particles do not interact individually with the charged leptons according to the Lagrangian \eqref{LCC'}. Instead, some {\it coherent} linear combinations of those massive states do. This is the essence of the problem: particle states of different masses cannot be coherently created or annihilated in quantum field theory. The transition amplitudes are summed together and interfere if and only if all external particle states are the same \cite{Bog-Shirk, Bjorken, Peskin, Coleman}. 

Technically, in QFT, the particle states $|\nu_1\rangle$ and $|\nu_2\rangle$ belong to distinct and orthogonal Hilbert spaces,  therefore any combination thereof can only be incoherent mixing. This is directly related to Haag's theorem \cite{Haag}. For more detailed considerations of this aspect in the context of neutrino mixing, see \cite{Tureanu, Kupiainen}.

Consequently, it is impossible to defend the claim that weak interactions produce coherent states like $c_1|\nu_1\rangle+c_2|\nu_2\rangle$, with $|c_1|^2+|c_2|^2=1$.

There are several other difficulties with the postulated formula \eqref{states_mix}, even if we choose to ignore the glaring fact that such states cannot be technically created in weak interactions:

\begin{itemize}
\item In order to work with the logic of a two-level system, the two sets of states, namely $(|\nu_e\rangle,|\nu_\mu\rangle)$ and $(|\nu_1\rangle,|\nu_2\rangle)$, have to be sets of eigenstates of two different Hamiltonians (see Appendix \ref{app_coherence}). Moreover, the states have to be unequivocally specified (for example, by their momentum or energy). While it is clear from \eqref{Sch_eq} that $(|\nu_1\rangle,|\nu_2\rangle)$ are Hamiltonian eigenstates, the flavour set $(|\nu_e\rangle,|\nu_\mu\rangle)$ does not fulfill this requirement.

\item The superposition on the r.h.s. involves states of particles of different masses, and in quantum mechanics as well as quantum field theory, particles of different masses represent different physical systems. The principle of superposition in quantum mechanics is valid for states of the same system, provided that no superselection rules are forbidding the coherent superposition. Recall that the principle of superposition is a consequence of the basic fact that a linear combinations of solutions of a differential equation (in this case, Schr\"odinger's equation) is also solution of the same equation. But the states $|\nu_1\rangle$ and $|\nu_2\rangle$ satisfy different differential equations, therefore the argument of superposition fails.

\item {In technical terms, if two particle states belong to different Hilbert spaces, then one can produce only probabilistic mixtures of those states, described by density matrices and in which the relative phase is arbitrary. Quantum superpositions with fixed relative coefficients are not allowed \cite{Mike-Ike}. Yet only quantum superposition can exhibit intereference (and oscillations), while the probabilistic mixtures cannot (see also \cite{Cozzella}). Let us recall that coherent superpositions of states are pure states, which correspond to vectors in the Hilbert space. Consequently, by the logic of coherent superposition, the flavour neutrino states should be vectors in a Hilbert space -- but such a Hilbert space does not exist.}

\end{itemize}

In short, formula \eqref{states_mix} with massive Fock states on the r.h.s. cannot be justified in either quantum field theory or quantum mechanics, although it is postulated to be valid in both. This is hardly any wonder, since this formula is simply a translation, in terms of states, of the rigorous formula \eqref{rotation} for field operators -- a procedure which is not meaningful in quantum field theory.

\section{The coherent scheme for neutrino oscillations}\label{coh sch}

We shall show that, in certain conditions, {\it formula \eqref{states_mix} can be derived with the required coherence, by changing the interpretation of flavour and massive states}. By those conditions, we ensure that both flavour and massive neutrinos can be treated as quantum mechanical states belonging to the same Hilbert space,  and well defined by association with specific Hamiltonians. 

The assumptions are the following:

\begin{enumerate}
\item The asymptotic states and the only physical states are massless neutrino states carrying flavour number. The Lagrangian is expressed in terms of the flavour fields (including also right-handed sterile ones), and the terms which are usually called mass terms, namely
\begin{eqnarray}\label{mass_terms}
- \left(\begin{array}{c c}
            \bar{\nu}_{eL}(x)&  \bar{\nu}_{\mu L}(x)
            \end{array}\right) \left(\begin{array}{c c}
            m_{ee} &m_{e\mu}\\
            m_{e\mu}&m_{\mu\mu}
            \end{array}\right)\left(\begin{array}{c}
            \nu_{e R}(x)\\
            {\nu_{\mu R}}(x)
            \end{array}\right)+h.c.,\nonumber
\end{eqnarray}
are treated as bilinear interaction terms. This assumption is in line with the general principle that all the masses of fundamental particles are due to some interaction and spontaneous gauge symmetry breaking, like the Brout--Englert--Higgs mechanism. In the case of Dirac masses, $m_{\ell\ell'}=v y_{\ell\ell'}/\sqrt2$, where $v$ is the Higgs field vacuum expectation value (vev) and $y_{\ell\ell'}$ stands for the matrix of Yukawa couplings that mix the neutrinos among themselves.

The massive states are only vacuum propagation states, and they do not interact weakly or in any other manner.

\item The massless electron neutrino and the muon neutrino of equal momenta are interpreted as degenerate states of a two-level system. The massive states of the same momentum are the eigenstates of the two-level system after the application of the perturbation.

\item In quantum mechanics, the Hilbert space of states of a given system is unique (Stone--von Neumann theorem), irrespective of the possible interactions that can be turned on or off. This cannot be reproduced rigorously with particles of different masses, which are born from orthogonal vacua. However, in the ultrarelativistic limit, the effects of the different vacua are negligible (see Appendix \ref{App_vac}). Consequently, we shall consider that massless and massive states are sharing the same Hilbert space and apply the formalism of a quantum mechanical two-level system. As we shall see later, the massless states represent particles (flavour neutrinos), while the massive states do not.

\item The number of particles is constant throughout the propagation and the particles' momentum remains constant. Even if the states switch from one flavour to the other, the change is viewed as a consequence of the interaction with a potential. This is ensured at quantum field theoretical level by the coherent scattering.

\end{enumerate}

\subsection{Potential associated with the bilinear mass terms}\label{Born}
\begin{figure}
     \centering
     \begin{subfigure}[b]{0.47\textwidth}
         \centering
         \includegraphics[width=\textwidth]{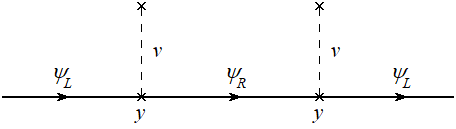}
         \caption{}
         \label{fig:y equals x}
     \end{subfigure}
     \hfill
     \begin{subfigure}[b]{0.47\textwidth}
         \centering
         \includegraphics[width=\textwidth]{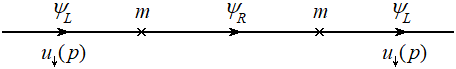}
         \caption{}
         \label{fig:three sin x}
     \end{subfigure}
        \caption{The interaction of the Weyl fields with the constant scalar field $v$ in (a) is equivalent to the mass insertions in (b).}
        \label{fig:Fig2}
\end{figure}

According to the above outlined programme, the weak interaction asymptotic states are massless flavour neutrinos. Once produced, they propagate in the "medium" created by the Higgs vev, which can be interpreted as an interaction with a constant background scalar field $v$ (see Fig. \ref{fig:Fig2}). The potential associated with the bilinear mass term changes the energy of the propagating massless particles.

We consider the simple Lagrangian:
\begin{eqnarray}\label{Lagr_Dirac}
{\cal L}&=&
{\cal L}_0+{\cal L}_{int},
\end{eqnarray}
where
\begin{eqnarray}
{\cal L}_0&=&\bar{\psi}_L(x)i\gamma^{\mu}\partial_{\mu}\psi_L(x)+\bar{\psi}_R(x)i\gamma^{\mu}\partial_{\mu}\psi_R(x)\cr
{\cal L}_{int}& =&- m\left(\bar{\psi}_L(x)\psi_R(x)+\bar{\psi}_R(x)\psi_L(x)\right).
\end{eqnarray}
Thus, we regard the "mass term" ${\cal L}_{int}$ as an interaction term of the Weyl spinors $\psi_L(x)$ and $\psi_R(x)$, which have the free-field expansions:
\begin{eqnarray}\label{Dirac_mode_exp}
\psi_L({\bf x},0)=\int\frac{d^3p}{(2\pi)^{3/2}\sqrt{2\tp}}\left(a_\da({\bf p})u_\da({\bf p})e^{i{\bf p\cdot x }}+b^\dagger_\ua({\bf p})v_\ua({\bf p})e^{-i{\bf p\cdot x }}\right),\nonumber\\
\psi_R({\bf x},0)=\int\frac{d^3p}{(2\pi)^{3/2}\sqrt{2\tp}}\left(a_\ua({\bf p})u_\ua({\bf p})e^{i{\bf p\cdot x }}+b^\dagger_\da({\bf p})v_\da({\bf p})e^{-i{\bf p\cdot x }}\right).
\end{eqnarray}
The corresponding Hamiltonian of interaction is:
\begin{eqnarray}\label{H_Dirac}
{\cal H}_{int}(x)= m\left(\bar{\psi}_L(x)\psi_R(x)+\bar{\psi}_R(x)\psi_L(x)\right) 
\end{eqnarray}
and the S-matrix reads:
\begin{eqnarray}S=\exp\left[iT\int d^4x{\cal H}_{int}(x)\right]
\end{eqnarray}

It is well known that one can convert the massless propagator into a massive propagator by summing the geometric series
\begin{eqnarray}
\frac{i}{\slashed p}+\frac{i}{\slashed p}(-im)\frac{i}{\slashed p}+\frac{i}{\slashed p}(-im)\frac{i}{\slashed p}(-im)\frac{i}{\slashed p}+\ldots=\frac{i}{\slashed p-m},
\end{eqnarray}
obtained by treating the mass term as an interaction terms, according to the separation in formula \eqref{Lagr_Dirac}. This inspires us to use a similar procedure for on-shell massless particles, as explained below.

We consider the lowest order nontrivial process depicted in Fig. \eqref{fig:Fig2}. Its Feynman amplitude is:
\begin{eqnarray}
i{\cal M}&=&-i m^2\bar u_\da({\bf p})P_R\frac{\slashed p}{p^2+i\epsilon}u_\da({\bf p})= -i\frac{m^2}{p^2+i\epsilon}\sum_\lambda\bar u_\lambda({\bf p})P_RP_R{\slashed p}P_Lu_\lambda({\bf p})\cr
&=&-i\frac{m^2}{p^2+i\epsilon}\mbox{Tr}\,\left[u_\lambda({\bf p})u_\lambda({\bf p})P_R{\slashed p}\right]=-i\frac{m^2}{p^2+i\epsilon}\mbox{Tr}\,\left[{\slashed p}P_R{\slashed p}\right] =- 2im^2.
\end{eqnarray}
The corresponding $T$-matrix element, where $S=1+iT$, is:
\begin{eqnarray}
\langle p|iT|p'\rangle=V (2\pi)\delta(E-E')\frac{1}{2VE}i{\cal M},\end{eqnarray}
where $V$ is the normalization volume.
Using the Born approximation formula from nonrelativistic quantum mechanics (see, for example, \cite{Peskin}), we can extract from this matrix element the potential energy $W$ of the corresponding interaction:
\begin{eqnarray}\label{Born_appr}
\langle p|iT|p'\rangle=- (2\pi)\delta(E-E')i {W}.\end{eqnarray}
Thus, we obtain:
\begin{eqnarray}\label{potential}
{W}= \frac{m^2}{E},\ \ \ \ E=\tp.\end{eqnarray}
We interpret this as the potential energy that the massless fermion of momentum $\bf p$ experiences during the propagation through the vacuum, which is seen as a constant, homogeneous medium. This potential has to be added to the kinetic energy $E=\tp$, to obtain the total energy during propagation,
\begin{eqnarray}\label{energy_m}
\tp+ \frac{m^2}{E},\ \ \ \ E=\tp.\end{eqnarray}

We note the discrepancy between \eqref{energy_m} and the energy of a relativistic particle of mass $m$, namely $\tp+ \frac{m^2}{2E}$.  This is due to the fact that the calculation above is done on the vacuum of massless particles, as explained in Appendix \ref{App_vac}. A genuine massive relativistic particle is not perfectly equivalent to a massless particle moving in a potential. 
Nevertheless, the factor $2$ is irrelevant -- all that matters is the dependence of the potential energy on the mass parameters (seen as "coupling constants") and the momentum of the particle. We shall return to this later in Sect. \ref{two-level}.

\subsection{Ultrarelativistic neutrinos as two-level oscillating system}\label{two-level}


\paragraph {\bf Flavour neutrinos as massless states of unperturbed Hamiltonian:} The left-helicity flavour neutrinos are produced in weak interactions as massless fermionic states. They are eigenstates of the free Hamiltonian
\begin{eqnarray}
H_{0}=\left(\begin{array}{c c}
\tp &
     0 \\
       0 &
      \tp
            \end{array}\right),
\end{eqnarray}
satisfying the free Schr\"odinger equation:
\begin{eqnarray}
i\partial_t\left(\begin{array}{c}
            |{\nu_e}({\bf p})\rangle\\
            |{\nu_{\mu}}({\bf p})\rangle
            \end{array}\right)= \left(\begin{array}{c c}
\tp &
     0 \\
       0 &
      \tp
            \end{array}\right)\left(\begin{array}{c}
            |{\nu_e}({\bf p})\rangle\\
            |{\nu_{\mu}}({\bf p})\rangle
            \end{array}\right).
\end{eqnarray}
This is the fundamental state of a degenerate two-level system. We omit to indicate the helicity of the states, but the convention is as in Standard Model, namely that all massless neutrinos are left-handed and all massless antineutrinos are right-handed.

%
%

\paragraph{\bf Bilinear mass terms as perturbation of the free Hamiltonian:} Once created as massless flavour states, the electron and muon neutrinos enter {\it suddenly} a "medium" with which they interact, similarly to the photons entering a dielectric. The difference is that the medium for neutrinos is a homogeneuous and constant scalar field $v$, which is normally called the vacuum expectation value of the Higgs field. The massless flavour neutrinos are coherently scattered in this medium, which does not get disturbed by their passage. The Lagrangian responsible for this part of the scheme is:
\begin{eqnarray}\label{bilin}
{\cal L}_0+{\cal L}_{mass}&=&\sum_{\ell=e,\mu}\left[\bar{\nu}_{\ell L}(x)i\slashed{\partial}\nu_{\ell L}(x) +\bar{\nu}_{\ell R}(x)i\slashed{\partial}\nu_{\ell R}(x)\right]\cr
&-& \left(\begin{array}{c c}
            \bar{\nu}_{eL}(x)&  \bar{\nu}_{\mu L}(x)
            \end{array}\right) \left(\begin{array}{c c}
            m_{ee} &m_{e\mu}\\
            m_{e\mu}&m_{\mu\mu}
            \end{array}\right)\left(\begin{array}{c}
            \nu_{e R}(x)\\
            {\nu_{\mu R}}(x)
            \end{array}\right)+h.c.
\end{eqnarray}
\begin{figure}
    \centering
    \begin{subfigure}[t]{0.4\textwidth}
        \centering
        \includegraphics[width=\linewidth]{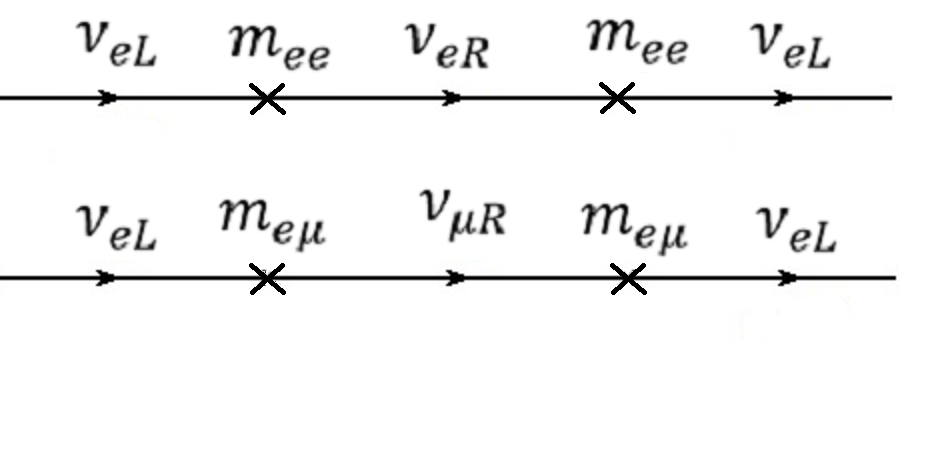} 
        \caption{$\nu_e\to\nu_e$} \label{fig:a}
    \end{subfigure}
    \hfill
    \begin{subfigure}[t]{0.4\textwidth}
        \centering
        \includegraphics[width=\linewidth]{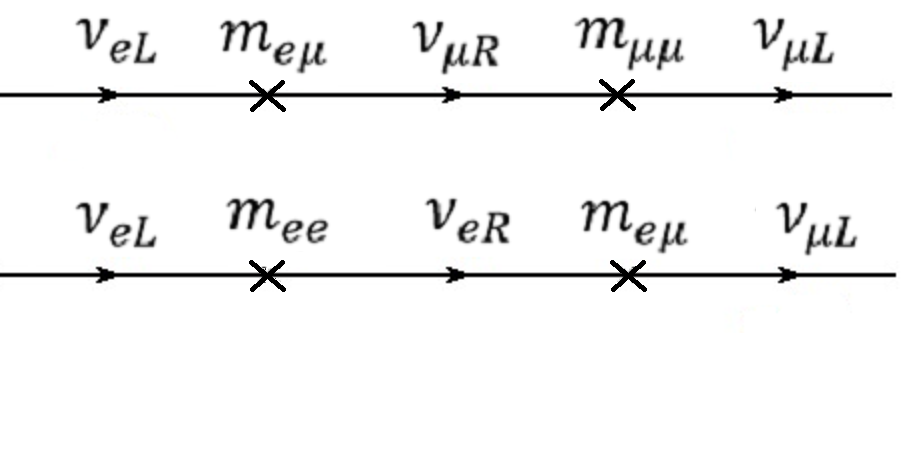} 
        \caption{$\nu_e\to\nu_\mu$} \label{fig:b}
    \end{subfigure}
%
    \begin{subfigure}[t]{0.4\textwidth}
        \centering
        \includegraphics[width=\linewidth]{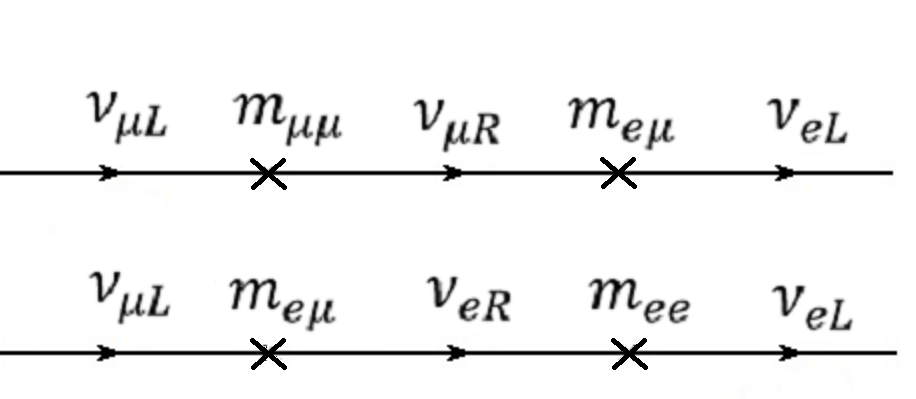} 
        \caption{$\nu_\mu\to\nu_e$} \label{fig:c}
    \end{subfigure}
    \hfill
    \begin{subfigure}[t]{0.4\textwidth}
        \centering
        \includegraphics[width=\linewidth]{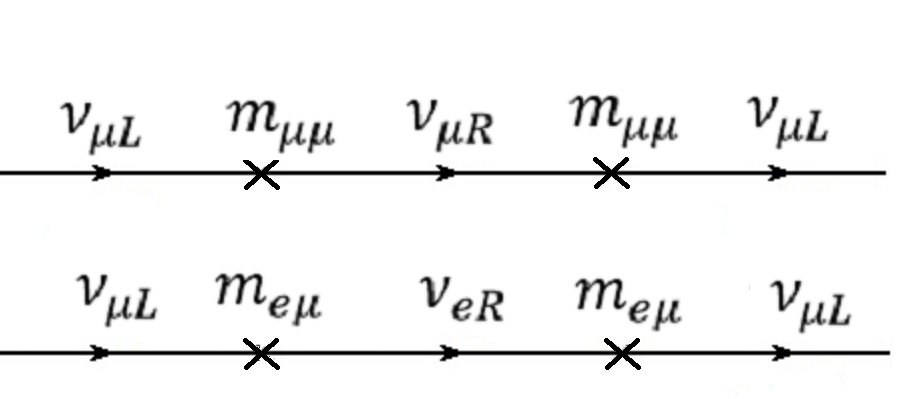} 
        \caption{$\nu_\mu\to\nu_\mu$} \label{fig:d}
    \end{subfigure}
\caption{Feynman diagrams for the bilinear interactions \eqref{bilin} in the lowest order of perturbation.}\label{fig:FD}
\end{figure}

We consider that the interaction with the vacuum produces a perturbation, in the form of a potential energy that mixes the two degenerate states and shifts as well the diagonal matrix elements of the Hamiltonian. The quantum mechanical interpretation as potential is obtained using the Born approximation, which relates the transition amplitude, calculated in quantum field theory, to the quantum mechanical potential scattering (see Sect. \ref{Born}). The Feynman amplitudes for the relevant processes described by the Feynman diagrams in Fig. \ref{fig:FD} are
\begin{eqnarray}
i{\cal M}_{\nu_e\to\nu_e}&=&-i (m_{ee}^2+m^2_{e\mu})\bar u_\da({\bf p})P_R\frac{\slashed p}{p^2+i\epsilon}u_\da({\bf p})=-2i (m_{ee}^2+m^2_{e\mu}),\cr 
i{\cal M}_{\nu_e\to\nu_\mu}&=&i{\cal M}_{\nu_\mu\to\nu_e}=-i m_{e\mu}(m_{ee}+m_{\mu\mu})\bar u_\da({\bf p})P_R\frac{\slashed p}{p^2+i\epsilon}u_\da({\bf p})=-2im_{e\mu}(m_{ee}+m_{\mu\mu}),\cr
 i{\cal M}_{\nu_\mu\to\nu_\mu}&=&-i (m_{\mu\mu}^2+m_{e\mu}^2)\bar u_\da({\bf p})P_R\frac{\slashed p}{p^2+i\epsilon}u_\da({\bf p})=-2i (m_{\mu\mu}^2+m_{e\mu}^2).
\end{eqnarray}
The corresponding potentials are collected in the interaction Hamiltonian:
\begin{eqnarray}
H_{int}=\left(\begin{array}{c c}
W_{11} &
     W_{12} \\
      W_{21} &
     W_{22}
            \end{array}\right)=
\frac{{ 1}}{\tp}\left(\begin{array}{c c}
m_{ee}^2+m^2_{e\mu}&
     m_{e\mu}(m_{ee}+m_{\mu\mu}) \\
      m_{e\mu}(m_{ee}+m_{\mu\mu}) &
      m_{\mu\mu}^2+m_{e\mu}^2
            \end{array}\right).
\end{eqnarray}
Thus, the total Hamiltonian of the massless flavour neutrinos in the physical vacuum becomes:
\begin{eqnarray}\label{total_H}
H=H_0+H_{int}= \left(\begin{array}{c c}
\tp &
     0 \\
       0 &
      \tp
            \end{array}\right)+\frac{{1}}{\tp}\left(\begin{array}{c c}
m_{ee}^2+m^2_{e\mu}&
     m_{e\mu}(m_{ee}+m_{\mu\mu}) \\
      m_{e\mu}(m_{ee}+m_{\mu\mu}) &
      m_{\mu\mu}^2+m_{e\mu}^2
            \end{array}\right),
\end{eqnarray}
leading to the perturbed Schr\"odinger's equations:
\begin{eqnarray}
i\partial_t\left(\begin{array}{c}
            |{\nu_e}({\bf p})\rangle\\
            |{\nu_{\mu}}({\bf p})\rangle
            \end{array}\right)= \frac{1}{\tp}\left(\begin{array}{c c}
\tp^2+(m_{ee}^2+m^2_{e\mu}) &
     m_{e\mu}(m_{ee}+m_{\mu\mu}) \\
       m_{e\mu}(m_{ee}+m_{\mu\mu}) &
      \tp^2+ ( m_{\mu\mu}^2+m_{e\mu}^2)
            \end{array}\right)\left(\begin{array}{c}
            |{\nu_e}({\bf p})\rangle\\
            |{\nu_{\mu}}({\bf p})\rangle
            \end{array}\right).
\end{eqnarray}

\paragraph{\bf Massive states as eigenstates of the perturbed Hamiltonian:} We assume that a left-helicity electron neutrino $|{\nu_e}({\bf p})\rangle$ is produced by weak interactions as massless eigenstate of $H_0$. Once produced, it enters the "medium" in which it interacts through the bilinear coupling.

%
This is equivalent to a {\it sudden switching on} of the bilinear interaction. Since the process is diabatic, the eigenstate $|{\nu_e}({\bf p})\rangle$ of $H_0$ starts to evolve according to the total Hamiltonian $H$, as a coherent superposition of the eigenstates of $H$. We diagonalize $H$ given by formula \eqref{total_H} using the transformation:
\begin{eqnarray}
U H U^T =\left(\begin{array}{c c}
            E_1 &0\\
            0&E_2
            \end{array}\right),\ \ \ \
U= \left(\begin{array}{c c}
            \cos\theta &\sin\theta\\
           -\sin\theta&\cos\theta
            \end{array}\right),
\end{eqnarray}
finding the eigenvalues:
\begin{eqnarray}\label{E_eigen}
E_{1,2}=\tp &+&\frac{1}{2\tp}\left[m_{ee}^2+m_{\mu\mu}^2+m_{e\mu}(m_{ee}+m_{\mu\mu})\right]\\\nonumber
&\mp&\frac{1}{2\tp}\sqrt{(m_{ee}^2-m_{\mu\mu}^2)^2+4m^2_{e\mu}(m_{ee}+m_{\mu\mu})^2}
\end{eqnarray}
and the mixing matrix:
\begin{eqnarray}\label{mixing angle}
U= \left(\begin{array}{c c}
            \cos\theta &\sin\theta\\
           -\sin\theta&\cos\theta
            \end{array}\right),\ \ \ \ \ \ \tan2\theta=\frac{2m_{e\mu}}{m_{\mu\mu}-m_{ee}}.
\end{eqnarray}
We denote
\begin{eqnarray}
m_{1,2}^2=\frac{1}{2}\left\{\left[m_{ee}^2+m_{\mu\mu}^2+m_{e\mu}(m_{ee}+m_{\mu\mu})\right]\mp\sqrt{(m_{ee}^2-m_{\mu\mu}^2)^2+4m^2_{e\mu}(m_{ee}+m_{\mu\mu})^2}\right\},
\end{eqnarray}
which gives
\begin{eqnarray}\label{masses'}
m_{1,2}&=&\frac{\sqrt 2}{2}\left\{\left(m_{ee}+m_{\mu\mu}\right)\mp\sqrt{(m_{ee}^2-m_{\mu\mu}^2)^2+4m^2_{e\mu}}\right\}.
\end{eqnarray}
According to \eqref{rotation_modes}, the eigenstates of $H_0$ are related to the eigenstates of $H$ by
\begin{eqnarray}\label{coherent_trans}
\left(\begin{array}{c}
            |{\nu_e}({\bf p})\rangle\\
            |{\nu_{\mu}}({\bf p})\rangle
            \end{array}\right)= \left(\begin{array}{c c}
            \cos\theta &\sin\theta\\
           -\sin\theta&\cos\theta
            \end{array}\right)\left(\begin{array}{c}
            |{\nu_1}({\bf p})\rangle\\
            |{\nu_2}({\bf p})\rangle
            \end{array}\right).
\end{eqnarray}

{Up to the constant $\sqrt{2}$, the formulas above coincide with the standard formulas \eqref{theta}--\eqref{masses}. The extra factor simply scales the parameters introduced in the Lagrangian, which are anyway free parameters, therefore apriorically unknown. For the analysis of experimental data, the rescaling of the mass parameters in the Lagrangian is irrelevant. 

From the point of view of the {\it quantum mechanical formalism},
\begin{itemize}
\item the transformation \eqref{coherent_trans} is a genuine change of basis in the Hilbert space of the two-level system, consequently it is {\it implicitly coherent}, as well as unitary;

\item both sets of states,  $|{\nu_e}({\bf p})\rangle, |{\nu_\mu}({\bf p})\rangle$ and $|{\nu_1}({\bf p})\rangle, |{\nu_2}({\bf p})\rangle$, are well-defined eigenstates of quantum mechanical Hamiltonians. 

\end{itemize}

From the point of view of {\it quantum field theory}, 

\begin{itemize}

\item the flavour states $|{\nu_e}({\bf p})\rangle$ and  $|{\nu_\mu}({\bf p})\rangle$ are massless Fock states and represent the asymptotic/free states of any process involving neutrinos;

\item the states $|{\nu_1}({\bf p})\rangle$ and  $|{\nu_2}({\bf p})\rangle$ are "propagation states" and not Fock states: they simply encode the information about the probability to find the initially produced flavour state or the other flavour state in the beam after a certain time, due to the mixing interaction.

\end{itemize}

}

\subsection{Oscillations and conversion in matter}

The coherent scheme described above integrated perfectly the propagation of neutrinos in matter. Assuming that matter is formed of first-generation particles, the coherent scattering of electron neutrinos on electrons by charged current weak interactions will be the dominating additional process compared to the vacuum propagation. This will add to the Hamiltonian \eqref{total_H}, as shown by Wolfenstein \cite{W}, an extra potential $V_{CC}=\sqrt 2 G_F n_e(x)$, where $n_e(x)$ is the density of electrons; equal contributions $V_{NC}$ from the coherent scattering of electron and muon neutrinos on neutrons have to be appropriately added as well, though their effect cancels in the oscillatory process. The Lagrangian of interaction (omitting the neutral current part) contains:
\begin{eqnarray}\label{Lmatter}
{\cal L}_{int}^{matter}=-\left(\begin{array}{c c}
            \bar{\nu_e}(x)&  \bar{\nu_{\mu}}(x)
            \end{array}\right) \left(\begin{array}{c c}
            m_{ee} &m_{e\mu}\\
            m_{e\mu}&m_{\mu\mu}
            \end{array}\right)\left(\begin{array}{c}
            {\nu_e}(x)\\
            {\nu_{\mu}}(x)
            \end{array}\right)-\left[\frac{g}{\sqrt2}\bar{\nu_e}_L(x)\gamma_{\mu} e_L(x)W^\mu+h.c.\right]\cr
\end{eqnarray}
and the coherent scattering processes are correspondingly:
\begin{eqnarray}\label{Feynman_matter}
\left(\begin{array}{c c}
\includegraphics[width=0.25\textwidth]{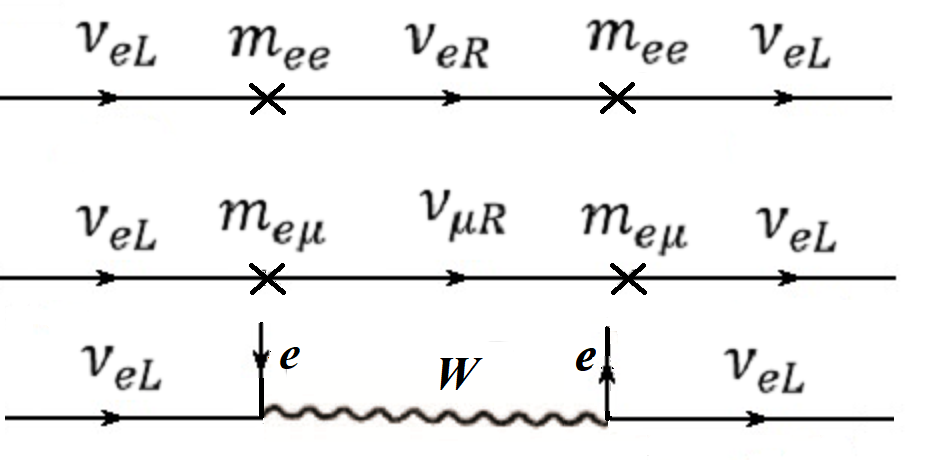}&
     \includegraphics[width=0.25\textwidth]{V12.png} \\
 & \\
     \includegraphics[width=0.25\textwidth]{V21.png} &
      \includegraphics[width=0.25\textwidth]{V22.png}
            \end{array}\right)
\end{eqnarray}
Using the procedure described in the previous section, we find the Hamiltonian of the two level system in matter:
\begin{eqnarray}\label{total_Hm}
H_{matter}&=& \left(\begin{array}{c c}
\tp+\frac{1}{\tp}[m_{ee}^2+m^2_{e\mu}]+V_{CC}&
     \frac{1}{\tp}[m_{e\mu}(m_{ee}+m_{\mu\mu}) ]\\
      \frac{1}{\tp}[m_{e\mu}(m_{ee}+m_{\mu\mu})] &
      \tp+\frac{1}{\tp}[m_{\mu\mu}^2+m_{e\mu}^2]
            \end{array}\right).
\end{eqnarray}
When we re-write the vacuum Hamiltonian\footnote{In the standard approach, the vacuum Hamiltonian \eqref{total_H'} is obtained by starting with the diagonalized Hamiltonian in the mass eigenstates basis and inverting the basis transformation \eqref{coherent_trans}. This procedure obliterates the physical meaning of the vacuum Hamiltonian in the flavour basis.} in terms of the parameters $m_1, m_2$ and $\theta$ given by \eqref{mixing angle} and \eqref{masses}, we find:
\begin{eqnarray}\label{total_H'}
H&=& {}\frac{m_1^2+m_2^2}{2E}\mathbf 1_{2\times 2}+\frac{1}{2E}\left(\begin{array}{c c}
-\Delta m^2\cos2\theta&
     \Delta m^2\sin2\theta\\
      \Delta m^2\sin2\theta &
      \Delta m^2\cos2\theta
            \end{array}\right).
\end{eqnarray}
Consequently,
\begin{eqnarray}\label{total_Hm}
H_{matter}&=& \left[\frac{m_1^2+m_2^2}{2E}+\frac{G_F}{\sqrt 2}n_e(x)\right]\mathbf 1_{2\times 2}\cr
&+&\frac{1}{2E}\left(\begin{array}{c c}
-\Delta m^2\cos2\theta+A_{CC}&
     \Delta m^2\sin2\theta\\
      \Delta m^2\sin2\theta &
      \Delta m^2\cos2\theta-A_{CC}
            \end{array}\right),
\end{eqnarray}
where $A_{CC}=2\sqrt 2 G_F n_e(x) E$.

This form of the Hamiltonian of flavour neutrinos in matter is the classical formula encountered in articles and books, barring a factor of 2 multiplying the energy $E$ in the denominator of the vacuum part.  

In our scheme, the only relevant bases in matter are the set of eigenstates of $H_0$ and the set of eigenstates of $H_{matter}$. The vacuum mass eigenstates play absolutely no role in the matter propagation, except for the obvious fact that they represent themselves instantaneous adiabatic states, for $n_e(x)=0$. This is in contrast with the standard treatment, where the interaction of the flavour neutrinos with the plasma is calculated using mass eigenstates in vacuum (see, for example, \cite{W} or \cite{giunti}), implying that at any time during the propagation in matter there are  {\it three bases} which are relevant, which is clearly one too many.
 
The present description justifies the assertion that the flavour neutrinos are always "the same", whether they are in vacuum or in matter, i.e.
\begin{eqnarray}\label{sameness}
|{\nu_e}({\bf p})\rangle&=& \cos\theta|{\nu_1}({\bf p})\rangle+ \sin\theta|{\nu_2}({\bf p})\rangle,\\
|{\nu_e}({\bf p})\rangle&=& \cos\theta_m|{\nu_{1m}}({\bf p})\rangle+ \sin\theta_m|{\nu_{2m}}({\bf p})\rangle,
\end{eqnarray}
where $|{\nu_{1m}}({\bf p})\rangle, |{\nu_{2m}}({\bf p})\rangle$ is the basis in which $H_{matter}$ is diagonal and $\theta_m$ is the corresponding mixing angle.

Thus, massless flavour neutrinos are the only interacting entities. When an electron neutrino is born in the core of the Sun, for example, it is emitted always as a massless neutrino, which enters suddenly a medium composed of the background constant scalar $v$ and a plasma, in which it interacts according to the Lagrangian \eqref{Lmatter}. As a result, it becomes subject to a series of coherent scattering microprocesses depicted in \eqref{Feynman_matter}, which slow down its passage and also mix the massless electron neutrino and muon neutrino states. Due to the sudden turning on of the interaction, the electron neutrino state starts to propagate as a coherent superposition of the adiabatic states $|{\nu_{1m}}({\bf p})\rangle, |{\nu_{2m}}({\bf p})\rangle$, with the coefficients depending on the density of electrons and the energy of the neutrino at the site of production. The propagation may be adiabatic, as in the MSW effect \cite{MS}, or not. Although the vacuum eigenstates do not feature at all in the present formalism for matter propagation, the expression \eqref{total_Hm} and all the calculations in terms of the set of vacuum parameters $(m_1,m_2,\theta)$ remains useful for comparison of data between vacuum and matter experiments/observations. The rest can be found in any textbook.

\section{Neutrino flavour nonconservation in weak interactions}\label{flavour_violation}

We shall turn now to the first requirement stated in the beginning, namely that the weak interactions of flavour neutrinos take place with flavour conservation.
Apparently, this requirement is a more serious hurdle. It is avowed times and again in books and articles as a fact, but to our knowledge it has never been quantitatively proven (see also Appendix \ref{app_flavour_violation}). Actually, quantum field theory leads us to expect the opposite, since the flavour violation in the quadratic part of the Lagrangian is likely to show up in any process, once quantum corrections are taken into account.

\begin{figure}
\centering
    \begin{subfigure}[t]{0.40\textwidth}
        \centering
        \includegraphics[width=\linewidth]{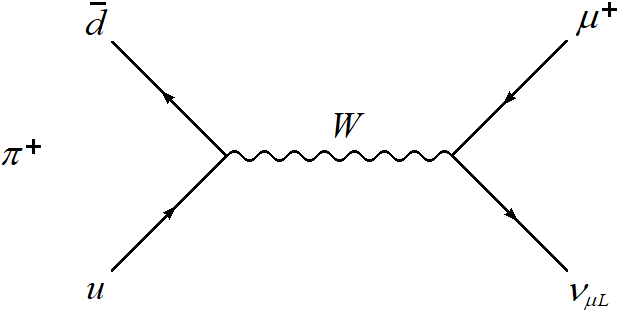} 
        \caption{$\pi^+\to\mu^++\nu_\mu$, tree level} \label{fig:pia}
    \end{subfigure}
\vskip 0.5cm
    \centering
    \begin{subfigure}[t]{0.45\textwidth}
        \centering
        \includegraphics[width=\linewidth]{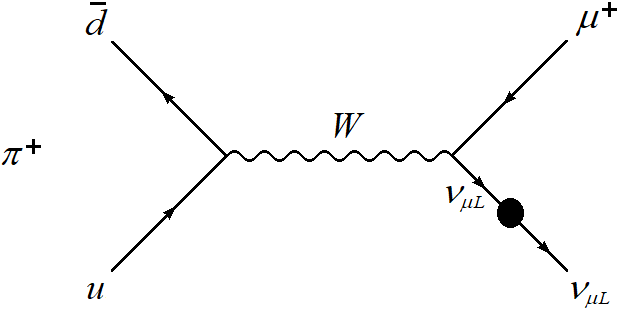} 
        \caption{$\pi^+\to\mu^++\nu_\mu$, renormalized} \label{fig:pib}
    \end{subfigure}
    \hfill
    \begin{subfigure}[t]{0.45\textwidth}
        \centering
        \includegraphics[width=\linewidth]{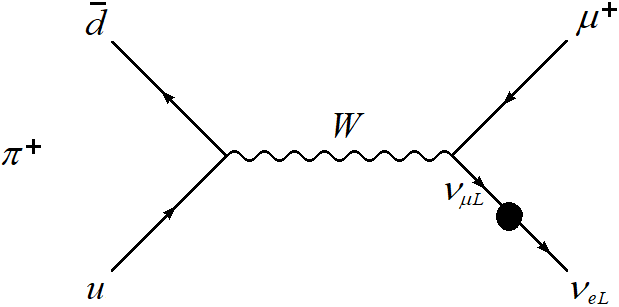} 
        \caption{$\pi^+\to\mu^++\nu_e$, renormalized} \label{fig:pic}
    \end{subfigure}
\caption{Feynman diagrams for the dominating pion decay to antimuons and neutrinos.}\label{fig:pion}
\end{figure}

For definiteness, let us consider the process of pion decay to antimuons. At tree level, we expect it to take place as described by Fig. \ref{fig:pia}, according to the charged current Lagrangian \eqref{LCC}. However, once we perform the external neutrino leg "renormalization", the situation changes drastically: since mixing processes such as those in \ref{fig:c} are allowed the pion decay to antimuons can lead to the production of either electron neutrinos, or muon neutrinos, as in Fig. \ref{fig:pib} and \ref{fig:pic}.  As a result, the inference of the number of a certain type of produced (or detected) flavour neutrinos by counting the charged leptons may be inaccurate, if the effect is substantial. The same flavour violation would happen in neutral current interactions.

\begin{figure}
\centering
\includegraphics[width=0.4\textwidth]{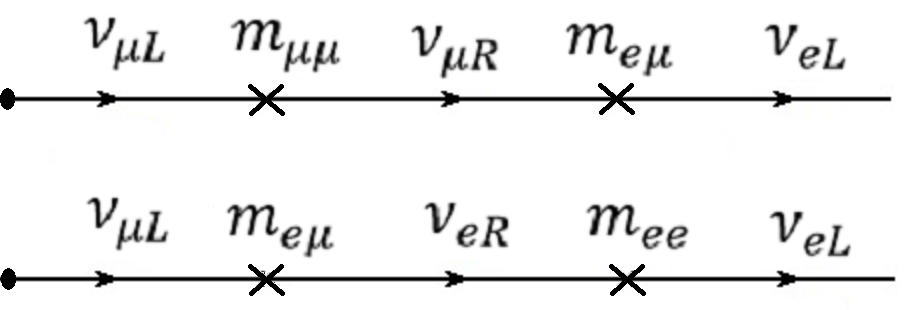} 
\caption{External neutrino line "renormalization" at the lowest order in perturbation.}\label{fig:external}
\end{figure}

According to Fig. \ref{fig:external}, in calculating the amplitude corresponding to Fig. \ref{fig:pic} in the second order in masses, we have to use for the external neutrino line
\begin{equation}
\frac{1}{E^2}m_{e\mu}\left(m_{\mu\mu}+m_{ee}\right)u({\bf p}),\ \ \ \ |{\bf p}|=E.
\end{equation}
When we square the amplitude, the probability of the pion decay with neutrino flavour violation in Fig. \ref{fig:pic} turns out to be of the order $m^4/E^4$. For ultrarelativistic neutrinos, this effect, which should manifest itself among other things as a "zero-length flavour conversion", is too small to be measured at the present level of experimental accuracy. We note also that the similar external leg dressing will appear also for neutrinos which preserve their flavour as in Fig. \ref{fig:pib}. The effects of the dressing are significant when the neutrino energies are very low, $E \sim m$. For this reason, in the previous sections we ignored the flavour violation in the weak interactions of high-energy neutrinos and adopted the view of a strictly sequential order of interactions. Nevertheless, the analysis of oscillations can be refined by adding the dressing as sketched above, though we expect relatively tiny contributions.

In Appendix \ref{app_flavour_violation} the flavour violation in the beta decay is calculated using the standard approach to neutrino oscillations, as described in Sect. \ref{intro}. The result is that, when using correctly the quantum field theory basic rules, even in the ultrarelativistic neutrino limit, the flavour violation effect is sizeable and at present measurable as zero-distance flavour conversion.

\section{Musing on the $K_0-\bar K_0$ system}\label{kaons}

Neutrino oscillations were modeled originally following the theory of $K_0-\bar K_0$ mixing and oscillations, proposed by Gell-Mann and Pais \cite{GMP} and Pais and Piccioni \cite{PP}. The problems of coherence  and "competition" between different interactions were clearly significant in these papers, though not much space is devoted to them. We can still glean the understanding of these issues in the interpretation of Gell-Mann, Pais and Piccioni.

\begin{figure}
     \centering
     \begin{subfigure}[b]{0.47\textwidth}
         \centering
         \includegraphics[width=\textwidth]{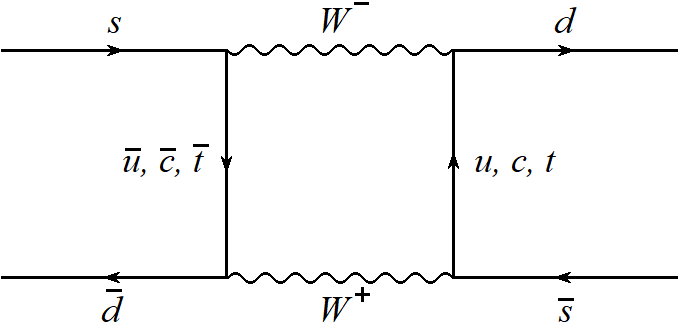}
         \caption{}
         \label{fig:y equals x}
     \end{subfigure}
     \hfill
     \begin{subfigure}[b]{0.47\textwidth}
         \centering
         \includegraphics[width=\textwidth]{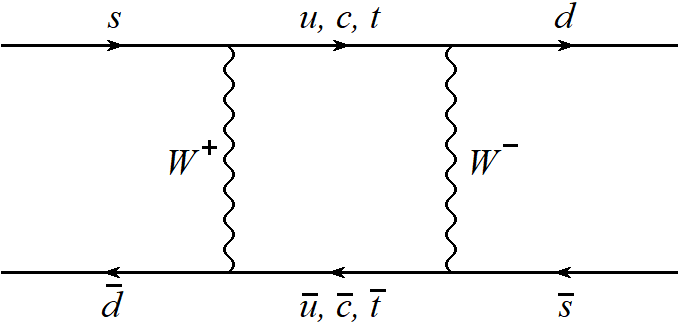}
         \caption{}
         \label{fig:three sin x}
     \end{subfigure}
        \caption{$K_0-\bar K_0$ mixing by weak interactions lead to the effective off-diagonal mass terms in \eqref{kaon_L}.}
        \label{fig:kaon}
\end{figure}

The bilinear part of the effective Lagrangian that describes the neutral kaon system is:
\begin{eqnarray}\label{kaon_L}
{\cal L}&=&\partial^\mu\Phi^\dagger(x)\partial_\mu\Phi(x)-\frac{1}{2}\left(\begin{array}{c c}
            \Phi^\dagger(x)&  \Phi(x)
            \end{array}\right) \left(\begin{array}{c c}
            M^2 &2\epsilon^2\\
            2\epsilon^2&M^2
            \end{array}\right)\left(\begin{array}{c}
            \Phi(x)\\
            \Phi^\dagger(x)
            \end{array}\right).
%
\end{eqnarray}
The off-diagonal mass parameter $\epsilon$ expresses effectively the weak interactions depicted in Fig. \ref{fig:kaon}, which mix the states $|K_0\rangle$ and $| \bar K_0\rangle$, created by strong interactions.

Gell-Mann and Pais make a point in \cite{GMP} that, if one has a free complex massive scalar field $\phi(x)$ of mass $M$, that field can be written in terms of its real and imaginary part fields as:
\begin{equation}\phi(x)=\frac{1}{\sqrt 2}(\phi_1(x)+\phi_2(x)),\end{equation}
with $\phi_1$ even and $\phi_2$ odd under ${\cal C}$-conjugation. The quanta of the real fields  $\phi_1$ and $\phi_2$ will be called $|K_1\rangle$ and $|K_2\rangle$ and the quanta of the complex field $\phi$ will be called $|K_0\rangle$ and $|\bar K_0\rangle$\footnote{In Ref. \cite{GMP} all these quantas are denoted by $\theta$, the name by which the $K$-mesons were called in 1955.}.Then, we have for the corresponding states with the same quantum numbers (same momenta and, if the field is not scalar, same spin polarization):
\begin{equation}\label{kaon_superpos}|K_1\rangle=\frac{1}{\sqrt 2}(|K_0\rangle+|\bar K_0\rangle).\end{equation}
In other words, the creation of a $|K_1\rangle$ quantum "corresponds physically to the creation, with equal
probability and with prescribed relative phase", of
either a $|K_0\rangle$ or a $|\bar K_0\rangle$.

This statement is not new, nor surprising. What is noteworthy is the acribia of Gell-Mann and Pais to make the statement, emphasizing that coherence ("prescribed relative phase") occurs when all the quanta $|K_0\rangle, |\bar K_0\rangle, |K_1\rangle, |K_2\rangle$ have {\it the same mass}. Later on in the same paper, they still rely on the formula \eqref{kaon_superpos} though the quanta $|K_1\rangle, |K_2\rangle$ are acknowledged to have different masses, "though the mass difference is surely tiny". This does not illuminate the point on how to preserve coherence when the two masses are different, but definitely suggests that their opinion was that the tininess of the mass difference with respect to the mass of either particle has a role in this game.

Soon after, in Ref. \cite{PP}, Pais and Piccioni proposed the experiment that bears their name to confirm the fact that the $|K_0\rangle$-meson is described by mixed states. In the same paper, they write for the first time the $K_0-\bar K_0$ oscillation formula. In the introduction of the paper, they return to the inverse of the formula \eqref{kaon_superpos},
\begin{equation}\label{kaon_superpos'}|K_0\rangle=\frac{1}{\sqrt 2}(|K_1\rangle+i|\bar K_2\rangle),\end{equation}
scrupulously mentioning again that  the superposition is "well-defined" (the equivalent of "prescribed relative phase" from \cite{GMP}), that all the particles are in the same momentum (and, if necessary, spin polarization) state, and adding the footnote: "This implies that we ignore the influence of the weak decay
interactions on the production" of the $|K_0\rangle$. In other words, the $|K_0\rangle$ state is produced strictly by strong interactions and the mixing terms proportional to $\epsilon^2$ in \eqref{kaon_L} are neglected, such that the masses of all states in \eqref{kaon_superpos'} are the same. The footnote has still another logical consequence: if the weak interactions are neglected at the production of $K_0$, they must spring in later on, during the propagation stage, leading up to the particles $K_1$ and $K_2$ acquiring the tiny mass difference that engenders the $K_0-\bar K_0$ oscillation. This is suggestive of a sequential treatement of the interactions, supported by the fact that the strangeness-violation effect in the production of  $K_0$ is truly negligible. 

Remark also that the conception of Gell-Mann, Pais and Piccioni was that the states $|K_0\rangle, |\bar K_0\rangle$ have definite mass and the formulas \eqref{kaon_superpos} and \eqref{kaon_superpos'} are written for well-defined momentum states. All these aspects of the original papers have survived to this day in the description of the neutral kaon systems, for which definite masses are assigned to $K_0$, $K_S$ and $K_L$ (the latter two being, barring CP violation effects, the same as $K_1$ and $K_2$, respectively).

In the manifestly coherent scheme presented in Sect. \ref{coh sch} for neutrinos, the (nonrelativistic) degenerate states $|K_0\rangle, |\bar K_0\rangle$ are described by the Hamiltonian:
\begin{eqnarray}\label{Ham_0_K}
H_{0}^{K}=\left(\begin{array}{c c}
M &
     0 \\
       0 &
      M
            \end{array}\right).
\end{eqnarray}
In quantum field theory, they are the eigenstates of the free Hamiltonian
 \begin{equation}{\cal H}_0=\partial^\mu\Phi^\dagger(x)\partial_\mu\Phi(x)-M^2\Phi^\dagger(x)\Phi(x).\end{equation}
The weak interaction sets in {\it after production} and mixes the two states due to the interaction Hamiltonian:
\begin{eqnarray}
{\cal H}_{int}^{K}=\epsilon^2\left(\Phi^\dagger(x)\Phi^\dagger(x)+\Phi(x)\Phi(x)\right),
\end{eqnarray}
such that 
\begin{eqnarray}
\langle \bar K_0({\bf p})|{\cal H}_{int}^{K}|K_0({\bf p})\rangle\neq 0.
\end{eqnarray}
The Feynman amplitude for the $|K_0({\bf p})\rangle\to |\bar K_0({\bf p})\rangle$ transition, in the lowest order, is:
\begin{equation}
i{\cal M}_{K_0\to \bar K_0}=i\epsilon^2.
\end{equation}
By Born approximation, we find the corresponding potential to be:
\begin{equation}
V_{K_0\to \bar K_0}=\frac{\epsilon^2}{2E}, \ \ \ E=M, \ \ \ \epsilon^2/M^2\ll 1.
\end{equation}
The Hamiltonian during propagation, omitting the absorptive part, is altered to
\begin{eqnarray}\label{Ham_0_K}
H^{K}=\left(\begin{array}{c c}
M &
     \frac{\epsilon^2}{2M} \\
       \frac{\epsilon^2}{2M} &
      M
            \end{array}\right)=\left(\begin{array}{c c}
\sqrt{M^2-\epsilon^2} &
     0 \\
       0 &
      \sqrt{M^2+\epsilon^2}
            \end{array}\right).
\end{eqnarray}
When the $K_0$ particle is nonrelativistic and the condition $\epsilon^2/M^2\ll 1$ is fulfilled, the eigenstates of $H_0$ and those of $H$ can be confidentely assumed to belong to the same Hilbert space (see Appendix \ref{App_vac}). 
The propagation eigenstates correspond to the weak-decay states $|K_1\rangle=|K_S\rangle$ and $|K_2\rangle=|K_L\rangle$ (CP violation ignored). The mixing is maximal and the coherence of the superposition is guaranteed by the two-level system formalism:
\begin{eqnarray}\label{K_superposition}
|K_0({\bf p})\rangle&=&\frac{1}{\sqrt{2}}\left(|K_1({\bf p})\rangle-|K_2({\bf p})\rangle\right),\cr
|\bar K_0({\bf p})\rangle&=&\frac{1}{\sqrt{2}}\left(|K_1({\bf p})\rangle+|K_2({\bf p})\rangle\right).
\end{eqnarray}
This scheme explains why the "prescribed relative phase" in \eqref{K_superposition} survives also when $|K_1\rangle$ and $|K_2\rangle$ have different masses, as long as the mass difference is very tiny compared to the mass of $K_0$.

\section{Outlook}

We have proposed a scheme, combining quantum field theory and formal aspects of quantum mechanics, in which neutrino oscillations are formulated in analogy with the quantum mechanical two-(or more) level systems. 
How we separate the Lagrangian into the free part and interaction part is a matter of choice (see \cite{NJL, Bog-Shirk}). The temptation is to diagonalize quickly whatever can be diagonalized. This is not always the most natural course of action. In the case of neutrinos, where the oscillation paradigm requires us to assume that flavour states are produced and detected, the sensible approach is to keep those states as the asymptotic/free ones. 
Consequently, the scheme gives a definite identity to the flavour states as massless Fock states. They have the leading role as asymptotic states in all processes involving neutrinos, be they weak interactions or mass-generating interactions. The latter set in suddenly after the high-energy flavour neutrinos are produced by weak interactions, and they create a "resistant" medium in which neutrinos propagate with flavour violation. 

The flavour states and the propagation (massive) states form two bases in the Hilbert space of the two-level system, as eigenstates of the unperturbed  Hamiltonian $H_0$ and the perturbed one, $H$. Nevertheless, from the quantum field theory point of view, only the massless states are Fock states. The coherence arises implicitly, technically following from the genuine change of basis between flavour and propagation states.

In this framework, vacuum oscillations and matter conversion of neutrinos arise both due to coherent scatterings of flavour neutrinos in media which mix the flavours, as shown in \eqref{Feynman_matter}.  The "sameness" of flavour neutrinos in vaccum and in matter is mathematically substantiated by their personal identity as massless eigenstates of the Hamiltonian $H_0$.

This treatment of neutrinos highlights the interactions which keep the flavours locked together during propagation and the transfer between flavours (see Fig. \ref{fig:FD}). Intuitively, the microscopic description of neutrino oscillations in vacuum is of a massless particle with a distance-dependent probability of being in one flavour state or another, determined by the "competition" between the flavour-preserving and flavour-violating interactions depicted in Fig. \ref{fig:FD}. The massive states are purely theoretical tools to determine, macroscopically, the flavour composition probability at a certain point after production. They do not feature in the calculations of neutrino production or detection probabilities. The mass parameters are coupling constants rather than kinematic parameters.  

The intuitive picture leads automatically to the conclusion that {\it any flavour neutrino produced in vacuum oscillates in perpetuity}, or at least until it is destroyed by weak interactions or other non-standard interactions. There is no chance that the propagation states $|\nu_1\rangle$ and $|\nu_2\rangle$ ever lose coherence, because they do not represent particle states and do not carry energy. Still, because neutrinos can travel undisturbed over huge spans of space, it is interesting to determine the velocity of propagation for an oscillating neutrino. Even if "in between two scatterings" the neutrinos propagate at the speed of light, the interaction itself is bound to slow them down\footnote{The situation is similar to the photon propagation in a dielectric: in between two scatterings, the photons propagate with the speed of light, however the change in phase between the incoming and scattered waves induces a perceived change in wavelength, therefore in speed.}.

Even if there is no decoherence when we speak about individual neutrinos, the propagation over long distances leads to averaging effects of the oscillations. We never know precisely where a given flavour neutrino was emitted by weak interaction, therefore the distance $L$ between production and detection sites is always affected by our ignorance. The energy resolution of the detector requires also averaging. As a result, in most of the experimental setups, the measurable probability of conversion is effectively constant and depending only on the elements of the mixing matrix (see, for example, \cite{giunti}). 

Here we have demonstrated the scheme on the simplest possible model, namely the two-flavour mixing with Dirac mass terms. Increasing the number of flavours or using Majorana mass terms (type II seesaw) should be straightforward. It will be interesting to study the case of mixing with sterile neutrinos which come with very heavy mass parameters (type I or III seesaw). 

Since this mechanism involves only massless neutrinos with definite helicity, it automatically follows that right-helicity neutrino states, or left-helicity antineutrino states can never be produced or observed in any frame. In contrast, in the standard approach using massive asymptotic states, there should be a sizeable amount of laboratory frame left-helicity nonrelativistic antineutrinos produced, for example, together with the electrons in the tail of the beta decay.

We expect the neutrinos with energies around and below $m_i, i=1,2$ given by \eqref{masses'} to be strongly influenced by the "dressing" discussed in Sect. \ref{flavour_violation}. Consequently, though a rigorous analysis in the present framework is still necessary, it is likely that the absolute values of the mass parameters be still experimentally accessible through the KATRIN experiment, for example. This scheme accommodates also the neutrinoless double beta decay \cite{Schechter-Valle} and the $2\nu$-mediated forces \cite{Bernabeu_2nu} as signaling the Majorana nature of the bilinear terms in the description of neutrinos, since the neutrino propagators will contain mass insertions. 

To summarize, we have presented a novel scheme for neutrino interactions and oscillations, in which the massless flavour neutrino states have the key role. The mechanism interpolates smoothly between high-energy neutrino phenomena and low-energy neutrino behaviour. The main achievement reported in this paper is the reconciliation of flavour neutrino production and detection with the coherence required for the description of oscillations in vacuum.

\subsection*{Acknowledgments}
I am grateful to Gabriela Barenboim, Jos\'e Bernab\'eu, Masud Chaichian, Jos\'e Gracia-Bond\'ia, Kazuo Fujikawa, Markku Oksanen and Adam Schwimmer for inspiring discussions, insightful comments and constructive suggestions.

\appendix

\section{Oscillation of states and coherence in two-level systems}\label{app_coherence}
The prototype for the description of neutrino oscillations is the quantum mechanical two-level system. This is a system with two stationary states, that get mixed by the sudden switching on of a time-independent interaction.

The system is initially described by Hamiltonian $H_0$ with the orthonormal basis states $|\psi_a\rangle$ and $|\psi_b\rangle$, with eigenvalues $E_a$ and $E_b$:
\begin{eqnarray}
H_0\left(\begin{array}{c}
            |\psi_a\rangle\\
            |\psi_b\rangle
            \end{array}\right)= \left(\begin{array}{c c}
E_a &
     0 \\
       0 &
      E_b
            \end{array}\right)\left(\begin{array}{c}
            |\psi_a\rangle\\
            |\psi_b\rangle
            \end{array}\right).
\end{eqnarray}
We assume that on can turn on an interaction, such that the system is described by the perturbed Hamiltonian 
$$H=H_0+ H_{int},$$ with
a new basis of stationary states $|\psi_1\rangle$ and $|\psi_2\rangle$, i.e. $$
H|\psi_i\rangle=E_i|\psi_i\rangle, \ \ i=1,2.
$$
In the general case, the perturbation shifts the diagonal elements of $H_0$ and it also mixes the states, i.e.
\begin{eqnarray}
H_{int}= \left(\begin{array}{c c}
W_{11} &
     W_{12} \\
      W_{21} &
     W_{22}
            \end{array}\right),\ \ \ \ \ W_{12} =
      W_{21}^*.
\end{eqnarray}
If $H_{int}$ is real, then $W_{12} = W_{21}=W$. According to the Stone--von Neumann theorem, all the representations of the canonical algebra for a given quantum mechanical system are equivalent, implying a {\it unitary change of basis} between the eigenstates of $H$ and the eigenstates of $H_0$:
\begin{eqnarray}\label{rotation_modes}
\left(\begin{array}{c}
           |\psi_a\rangle\\
           |\psi_b\rangle
            \end{array}\right)= \left(\begin{array}{c c}
            \cos\theta &\sin\theta\\
           -\sin\theta&\cos\theta
            \end{array}\right)\left(\begin{array}{c}
            |\psi_1\rangle\\
            |\psi_2\rangle
            \end{array}\right).
\end{eqnarray}
In other words, the states $|\psi_{a}\rangle$ and $|\psi_{b}\rangle$ are coherent superpositions of the states $|\psi_{1}\rangle$ and $|\psi_{2}\rangle$.

The same rotation matrix has to diagonalize the Hamiltonian $H$  by a similarity transformation:
\begin{eqnarray}
UH U^\dagger= \left(\begin{array}{c c}
E_{1} &
     0 \\
      0 &
     E_2
            \end{array}\right),\ \ \ \ U=\left(\begin{array}{c c}
            \cos\theta &\sin\theta\\
           -\sin\theta&\cos\theta
            \end{array}\right), \ \ \ \ \tan 2\theta=\frac{2W}{E_a-E_b+W_{22}-W_{11}}.
\end{eqnarray}
The eigenvalues of $H$ are
\begin{equation}E_{1,2}=\frac{1}{2}(E_a+E_b+W_{11}+W_{22})\mp\frac{1}{2}\sqrt{(E_a-E_b+W_{11}-W_{22})^2+4W^2}.\end{equation}

In the treatment of two-level system, it is customary to write the Hamiltonian \eqref{total_H} as
\begin{eqnarray}\label{H_Rabi}
H=H_0+H_{int}= \frac{1}{2}\left(\begin{array}{c c}
-\Delta&
     \Omega \\
      \Omega &
      \Delta
            \end{array}\right)+E_0\mathbf{1}_{2\times 2},
\end{eqnarray}
where
\begin{eqnarray}\label{E_0}
E_0=\frac{1}{2}(E_a+E_b+W_{11}+W_{22}).\end{eqnarray}
The variable $\Delta=E_a-E_b+W_{11}-W_{22}$ is called the {\it detuning} and $\Omega =2W$ is called the {\it Rabi frequency} of the system, in remembrance of two-level atoms coupled with an electromagnetic field which couples the two levels.
The mixing angle is then given by
\begin{eqnarray}
\tan2\theta=\frac{\Omega}{\Delta},
\end{eqnarray}
which shows that resonance is obtained for zero detuning. Both $\Delta$ and $\Omega$ may depend on time, in which case the transitions between levels can be adiabatic or not. The eigenvalues \eqref{E_eigen} of the Hamiltonian $H$ are:
\begin{eqnarray}\label{E_eigen'}
E_{1,2}=E_0\mp\frac{1}{2}\sqrt{\Delta^2+\Omega^2}.
\end{eqnarray}

We consider a time-independent perturbation. The system is prepared in the stationary state $|\psi_a\rangle$ and it evolves with $H_0$. At 
$t=t_0$, the interaction is turned on {\it suddenly (diabatically)}, such that the system does not transition slowly into a stationary state of $H$, but it remains for an instant in the state $|\psi_a\rangle=|\psi(t_0)\rangle$. Since the initial state $|\psi_a\rangle$ is a {\it coherent superposition} of the states $|\psi_1\rangle$ and $|\psi_2\rangle$, after evolving with the Hamiltonian $H$ for a period of time $\Delta t$, the system will be in the state
\begin{equation}|\psi (t)\rangle=\cos\alpha e^{-iE_1\Delta t}|\psi_1\rangle+\sin\alpha e^{-iE_2\Delta t}|\psi_2\rangle.\end{equation}
At the time $t=t_0+\Delta t$ we remove suddenly the interaction and determine the state of the system, which can be either of the $H_0$ eigenstates, $|\psi_a\rangle$ or $|\psi_b\rangle$. The probability that the system has transitioned into the state $|\psi_b\rangle$ is:
\begin{equation}{\cal P}_{|\psi_a\rangle\to|\psi_b\rangle}=|\langle \psi_b|e^{-iH\Delta t}|\psi_a\rangle|^2\sim \sin^2\left(\frac{\Delta E}{2}\Delta t\right),\ \ \ \ \ \Delta E= E_2-E_1=\sqrt{\Delta^2+\Omega^2} .\end{equation}

The question is whether the above simple picture can be applied to particle oscillations. Let us note the following facts regarding the two-level system:
i) the states of the two bases are well defined as stationary states of either $H_0$ or $H$; ii) the coherent superposition of states (leading to interference and finally to oscillation) is achieved by {\it turning on/off suddenly the interaction.}


\section{Particles of different masses and their Hilbert spaces}\label{App_vac}

\subsection{Spinors}

Consider $a_{\lambda}({\bf p})$ and $b_{\lambda}({\bf p})$ annihilation operators for free spinor particles and antiparticles of mass $M$ and helicity $\lambda$, operating on a vacuum $|0\rangle$, and  $A_{\lambda}({\bf p})$ and $B_{\lambda}({\bf p})$ the corresponding operators for massive spinor particles of mass $M+m$, operating on the vacuum $|\Phi\rangle$. All the operators satisfy canonical anti-commutation relations. The two sets of operators are connected by the Bogoliubov transformations:
\begin{eqnarray}\label{BT}
A_{\lambda}({\bf p})&=&\alpha_{\tp}a_{\lambda}({\bf p})+\beta_{\tp}b^\dagger_{\lambda}(-{\bf p}),\cr
B_{\lambda}({\bf p})&=&\alpha_{\tp}b_{\lambda}({\bf p})-\beta_{\tp}a^\dagger_{\lambda}(-{\bf p}),\ \ \
\end{eqnarray}
where 
\begin{eqnarray}\label{BT_coeff}
\alpha_{\tp}=\sqrt{\frac{\Omega_\tp+\omega_\tp}{2\Omega_\tp}},\ \
\beta_{\tp}=\sgn\,\lambda\,\sqrt{\frac{\Omega_\tp-\omega_\tp}{2\Omega_\tp}},\ \ \ |\alpha_{\tp}|^2+|\beta_{\tp}|^2=1.
\end{eqnarray}
and
\begin{eqnarray}\label{Omega}
\omega_{\tp}=\sqrt{{\tp}^2+M^2},\ \ \ \Omega_{\tp}=\sqrt{{\tp}^2+(M+m)^2}.
\end{eqnarray}
The vacua $|0\rangle$ and $|\Phi\rangle$ are orthogonal in the infinite momentum and infinite volume limit,
$$\langle 0|\Phi\rangle=0,$$
therefore the Fock spaces built on either of these two vacua are orthogonal as well. In other words, a particle state of a certain mass cannot be written as a linear superposition of particle states of different mass(es).

If we take $M=0$, then
\begin{eqnarray}\label{BT_coeff_massless}
\alpha_{\tp}=\sqrt{\frac{1}{2}\left(1+\frac{\tp}{\Omega_{\tp}}\right)},\ \
\beta_{\tp}=\sgn\,\lambda\,\sqrt{\frac{1}{2}\left(1-\frac{\tp}{\Omega_{\tp}}\right)}.
\end{eqnarray}
Now, $A^\dagger_{\lambda}({\bf p})|\Phi\rangle$ represents a one-particle state with mass $m$ and $\alpha_{\tp}a^\dagger_{\lambda}({\bf p})|0\rangle$ represents a massless one-particle state. They are rigorously orthogonal, so we cannot really compare them. But our purpose is to view the massive particle as a massless one, with the mass contribution to its energy given by a potential. The closest we can get to this interpretation in quantum field theory is to compare the states $A^\dagger_{\lambda}({\bf p})|0\rangle$ and $a^\dagger_{\lambda}({\bf p})|0\rangle$. Although $A^\dagger_{\lambda}({\bf p})|0\rangle$ is genuinely massless, it still carries the memory of the free massive Hamiltonian $H$.  

Applying the hermitian conjugate of \eqref{BT} on the massless vacuum $|0\rangle$, we obtain:
\begin{equation}
A^\dagger_{\lambda}({\bf p})|0\rangle=\alpha_{\tp}a^\dagger_{\lambda}({\bf p})|0\rangle.
\end{equation}
In the limit $m/\tp\ll 1$, we have
\begin{equation}
A^\dagger_{\lambda}({\bf p})|0\rangle=\left( 1-\frac{1}{8}\frac{m^2}{\tp^2}\right) a^\dagger_{\lambda}({\bf p})|0\rangle,
\end{equation}
therefore the two states coincide up to the order $m/\tp$. The difference of the order ${\cal O}(m^2/\tp^2)$ indicates that the energy of the massless state in a potential, $A^\dagger_{\lambda}({\bf p})|0\rangle$, will be different from the energy of the bare massless state $a^\dagger_{\lambda}({\bf p})|0\rangle$. In  Sect. \ref{Born} we confirm that the energy difference is of this order by using a Feynman diagram approach followed by Born approximation.

\subsection{Scalars}
We consider $a({\bf p})$ and $b({\bf p})$ the canonical annihilation operators for free scalar particles and antiparticles of square mass $M^2$, operating on a vacuum $|0\rangle$, and  $A({\bf p})$ and $B({\bf p})$ the corresponding operators for scalar particles of square mass $M^2+\epsilon^2$, operating on the vacuum $|\Phi\rangle$. The two sets of operators are connected by the Bogoliubov transformations:
\begin{eqnarray}\label{BT_scalar}
A({\bf p})&=&\alpha_{\tp}a({\bf p})+\beta_{\tp}b^\dagger(-{\bf p}),\cr
B({\bf p})&=&\alpha_{\tp}b({\bf p})-\beta_{\tp}a^\dagger(-{\bf p}),
\end{eqnarray}
where 
\begin{eqnarray}\label{BT_coeff_scalar}
\alpha_{\tp}=\frac{\Omega_\tp+\omega_\tp}{2\sqrt{\omega_\tp\Omega_\tp}},\ \ \ \ 
\beta_{\tp}=\frac{\Omega_\tp-\omega_\tp}{2\sqrt{\omega_\tp\Omega_\tp}},\ \ \ \ \ \ |\alpha_{\tp}|^2-|\beta_{\tp}|^2=1.
\end{eqnarray}
and
\begin{eqnarray}\label{Omega_scalar}
\omega_{\tp}=\sqrt{{\tp}^2+M^2},\ \ \ \Omega_{\tp}=\sqrt{{\tp}^2+(M^2+\epsilon^2)}.
\end{eqnarray}
As before, we apply the operatorial equation \eqref{BT_scalar} to the vacuum of the particles with mass $M$:
\begin{equation}
A^\dagger({\bf p})|0\rangle=\alpha_{\tp}a^\dagger({\bf p})|0\rangle.
\end{equation}
In the limit $\tp/M\ll1,\ \epsilon^2/M^2\ll 1$,
\begin{equation}
A^\dagger({\bf p})|0\rangle=\left( 1-\frac{1}{16}\frac{\epsilon^4}{M^4}\right) a^\dagger({\bf p})|0\rangle.
\end{equation}

\section{Comments on the flavour violation in the weak interactions of neutrinos}\label{app_flavour_violation}

Here we show that, in the standard approach to oscillations, in which the flavour states are defined exclusively through the superposition \eqref{states_mix}, the flavour violation in weak interactions is actually quite significant, even for ultrarelativistic neutrinos.

Consider the production of antineutrinos by beta decay:
$$n\to p+\bar\nu_\ell+e^-.$$
The matrix element is given by:
\begin{eqnarray}
\langle \bar\nu_\ell, e^-|\bar e(x)\gamma^\alpha(1-\gamma_5)\nu_e(x)|0\rangle  C_\alpha(n,p,\ldots),
\end{eqnarray}
where $C_\alpha(n,p,...)$ represents the contributions of the neutron and proton (depending on their fixed momenta and energies, as well as on the parameters of the leptons), including the coupling constant.
To assess the flavour violation, we have to consider $|\nu_\ell\rangle=|\nu_\mu\rangle$. The spirit of the traditional approach is that the mass eigenstates are produced and annihilated in the weak processes, therefore one has to use \eqref{rotation} and \eqref{states_mix} to re-express the amplitude:
\begin{eqnarray}\label{amplit}
&&\langle\bar\nu_\mu, e^-|\bar e(x)\gamma^\alpha(1-\gamma_5)\nu_e(x)|0\rangle  C_\alpha(n,p,\ldots)\cr
&&=-\sin\theta\cos\theta\, C_\alpha(n,p,m_1)\langle \bar\nu_1,e^-|\bar e(x)\gamma^\alpha(1-\gamma_5)\nu_1(x)|0\rangle\cr
&&+\sin\theta\cos\theta\, C_\alpha(n,p,m_2)\langle \bar\nu_2,e^-|\bar e(x)\gamma^\alpha(1-\gamma_5)\nu_2(x)|0\rangle.
\end{eqnarray}
We do not use momentum identifiers for the states because the conservation of energy and momentum would make it cumbersome. Anyway, it is assumed that the initial state is well defined kinematically. The momenta of the final states are a source of confusion.
This formula is customarily used to justify that flavour violation is practically nill in weak interaction. The argument is based on taking the ultrarelativistic limit in the amplitude formula \eqref{amplit}, in which case the factors depending on the masses of the antineutrinos become vanishingly small and \eqref{amplit} becomes zero because of the opposite sign in the two terms. However, this argument is misleading.

Formula \eqref{amplit} expresses the sum of two amplitudes in which the final particles differ. According to the rules of quantum field theory, {\it the two amplitudes do not interfere (their relative phase is irrelevant) and, in order to calculate the transition probability, we have to square each amplitude individually and add the results}, namely:
\begin{eqnarray}\label{probab}
{\cal P}_{n\to p+\bar\nu_\mu+e^-}&=&\frac{1}{4}\sin^22\theta\, |C_\alpha(n,p,m_1)|^2|\langle \bar\nu_1,e^-|\bar e(x)\gamma^\alpha(1-\gamma_5)\nu_1(x)|0\rangle|^2\cr
&+&\frac{1}{4}\sin^22\theta\, |C_\alpha(n,p,m_2)|^2|\langle \bar\nu_2,e^-|\bar e(x)\gamma^\alpha(1-\gamma_5)\nu_2(x)|0\rangle|^2.
\end{eqnarray}
The ultrarelativistic limit has to be taken in the probability formula \eqref{probab}. Neglecting all the powers of $m_i/\tp,\  i=1,2$, where $\tp$ is the momentum of the antineutrino, we obtain:
\begin{eqnarray}\label{probab_violation}
{\cal P}_{n\to p+\bar\nu_\mu+e^-}&=&\frac{1}{2}\sin^22\theta\,{\cal P}^{SM}_{n\to p+\bar\nu_e+e^-},
\end{eqnarray}
where ${\cal P}^{SM}_{n\to p+\bar\nu_e+e^-}$ is the Standard Model probability of the flavour-conserving reaction beta decay with the same kinematic configuration. Thus we find a sizable probability of ultrarelativistic muon antineutrino production in the beta decay, and this probability does not depend on the ratio between the mass and the momentum of neutrinos. It should consequently give a measurable "zero-length conversion" probability, which actually has never been observed experimentally.

The situation is actually even more implausible, because, if we calculate in the same manner as above the probability of the beta decay with electron antineutrinos in the ultrarelativistic limit, we find the expected result:
\begin{eqnarray}\label{probab_conserv}
{\cal P}_{n\to p+\bar\nu_e+e^-}&=&{\cal P}^{SM}_{n\to p+\bar\nu_e+e^-}.
\end{eqnarray}
Putting together \eqref{probab_violation} and \eqref{probab_conserv} we come to the conclusion that, by taking into account the mixing, the probability of neutron decay into proton, electron and antineutrino {\it increases} compared to the same probability calculated in the Standard Model (with zero neutrino masses). The bizarre result is due to the fact that the apparent orthogonality of the states $|\nu_e\rangle$ and $|\nu_\mu\rangle$ in \eqref{states_mix} is irrelevant in the decay probability computation: the flavour neutrino state could have been $|\nu_\mu\rangle=e^{i\delta}\sin\theta|\nu_1\rangle+\cos\theta|\nu_2\rangle$, with arbitrary real $\delta$, and the result would have been the same. This reflects the fact that coherence between states with different masses is an impossibility in quantum field theory. 

Assuming that we consider the mass eigenstates as asymptotic states, how should one calculate the beta decay probability in a sensible manner? The answer is straightforward: by using the Lagrangian \eqref{LCC'} and the principles of quantum field theory (see also \cite{Shrock_PLB, Shrock_PRD}, where the {\it incoherent decay} of pions into different massive neutrino states is acknowledged and exploited for experimental signatures). According to the Lagrangian \eqref{LCC'}, two independent processes contribute to the beta decay:
\begin{eqnarray}
n\to p+\bar\nu_1+e^-\cr
n\to p+\bar\nu_2+e^-.
\end{eqnarray}
With the previous notations, their transition amplitudes are:
\begin{eqnarray}\label{amplit_sensible}
&&\langle\bar\nu_1, e^-({\bf k})|\bar e(x)\gamma^\alpha(1-\gamma_5)\nu_e(x)|0\rangle  C_\alpha(n,p,\ldots)\cr
&&=\cos\theta\, C_\alpha(n,p,m_1)\langle \bar\nu_1,e^-({\bf k})|\bar e(x)\gamma^\alpha(1-\gamma_5)\nu_1(x)|0\rangle\cr
&&\langle\bar\nu_2, e^-({\bf k})|\bar e(x)\gamma^\alpha(1-\gamma_5)\nu_e(x)|0\rangle  C_\alpha(n,p,\ldots)\cr
&&=\sin\theta\, C_\alpha(n,p,m_2)\langle \bar\nu_2,e^-({\bf k})|\bar e(x)\gamma^\alpha(1-\gamma_5)\nu_2(x)|0\rangle,
\end{eqnarray}
and the probabilities:
\begin{eqnarray}\label{probab_sensible}
{\cal P}_{n\to p+\bar\nu_1+e^-}&=&\cos^2\theta\, |C_\alpha(n,p,m_1)|^2|\langle \bar\nu_1,e^-({\bf k})|\bar e(x)\gamma^\alpha(1-\gamma_5)\nu_1(x)|0\rangle|^2\cr
{\cal P}_{n\to p+\bar\nu_2+e^-}&=&\sin^2\theta\, |C_\alpha(n,p,m_2)|^2|\langle \bar\nu_2,e^-({\bf k})|\bar e(x)\gamma^\alpha(1-\gamma_5)\nu_2(x)|0\rangle|^2.
\end{eqnarray}
In the ultrarelativistic antineutrino limit, neglecting all the powers of $m_i/\tp$, we obtain:
\begin{eqnarray}
{\cal P}_{n\to p+{antineutrinos}+e^-}&=&{\cal P}^{SM}_{n\to p+\bar\nu_e+e^-},
\end{eqnarray}
which is a sound result. Nevertheless, in this calculation flavour neutrino states do not enter in any way.

\end{document}